\documentclass[runningheads]{llncs}

\usepackage{amsmath,amssymb,amsfonts}
\usepackage{algorithmic}
\usepackage{graphicx}
\usepackage{textcomp}
\def\BibTeX{{\rm B\kern-.05em{\sc i\kern-.025em b}\kern-.08em
    T\kern-.1667em\lower.7ex\hbox{E}\kern-.125emX}}

\usepackage{color}

\usepackage{url}

\usepackage[numbers]{natbib}

\begin{document}

\title{Sealed Computation: \\ Abstract Requirements for Mechanisms to Support Trustworthy Cloud Computing}

\author{%
Lamya	Abdullah\inst{1,2}\and 
Felix	Freiling\inst{1}	\and
Juan	Quintero\inst{1,2}	\and
Zinaida	Benenson\inst{1}	
}%

\institute{
Friedrich-Alexander-Universit\"at Erlangen-N\"urnberg (FAU), Erlangen, Germany \\
felix.freiling@cs.fau.de, zinaida.benenson@fau.de.
\and
Uniscon GmbH, Munich, Germany \\
lamya.abdullah@uniscon.de, juan.quintero@uniscon.de.
}

\maketitle

\begin{abstract}
  In cloud computing, data processing is delegated to a remote party
  for efficiency and flexibility reasons. A practical user requirement
  usually is that the confidentiality and integrity of data processing
  needs to be protected. In the common scenarios of cloud computing
  today, this can only be achieved by assuming that the remote party
  does not in any form act maliciously. In this paper, we propose an
  approach that avoids having to trust a single entity.  Our approach
  is based on two concepts: (1) the technical abstraction of
  \emph{sealed computation}, i.e., a technical mechanism to confine
  the processing of data within a tamper-proof hardware container, and
  (2) the additional role of an auditing party that itself cannot add
  functionality to the system but is able to check whether the system
  (including the mechanism for sealed computation) works as
  expected. We discuss the abstract technical and procedural
  requirements of these concepts and explain how they can be applied
  in practice.
\end{abstract}


\section{Introduction} 
\label{sec:intro}

Cloud computing has become widespread as it allows for supplying and
utilizing computation resources in an on-demand fashion. This reduces
cost, increases flexibility and improves infrastructure scalability \cite{Mell2009:effectivelyUsingCloud}. Cloud computing is increasingly being adapted for services provided by networks of small devices,
commonly referred to as the \emph{Internet of Things} (IoT). IoT Cloud \citep{alam2010senaas} or ``Cloud of Things'' (CoT) \citep{aazam2014cloud} 
  provides resources such as storage, analytics tools and shared configurable computing resources to reduce the cost and complexity associated with the IoT systems.

When the data processing and storage are delegated to a cloud
provider, users of cloud services usually have to trust the cloud
provider to act as expected. However, in common cloud deployments,
there is no technical guarantee that a single malicious insider like a
system administrator or a person with physical access to the cloud
infrastructure does not tamper with code and data. Hence cloud clients should be provided some technical
guarantees and indications that the cloud service is trustworthy.

As an example, consider the scenario of an IoT Cloud implementation for \emph{usage-based insurance}
(UBI) \citep{karapiperis2015cipr}, a novel car
insurance business model, where the insurance company calculates
premiums based on drivers’ behavior using actual driving data.
 
In UBI,
participating cars are equipped with a telematics devices to collect
driving data such as location, speed, acceleration, cornering, and
other details. Driving data are processed to get a 
ranking based on personal driving behavior. Using the driver ranking,
the insurance company calculates a customized premium to the
policyholder employing a more accurate risk estimate, reducing
incurred losses \citep{derikx2016can,soleymanian2017sensor} and
offering a bonus in the case of good driving behavior. 

UBI promises many benefits such as, for the insurance companies,
reducing incurred losses through accurate risk estimates
\citep{soleymanian2017sensor,derikx2016can} and, for the
policyholders (drivers), improving their driving style through feedback and
decreasing their premiums.  But
obviously, UBI also raises concerns, such as user discrimination 
\citep{karapiperis2015cipr}, and consumer privacy \citep{soleymanian2017sensor,derikx2016can}.

\begin{figure} [tbh]
 \centering
  \includegraphics[width=0.95\textwidth]{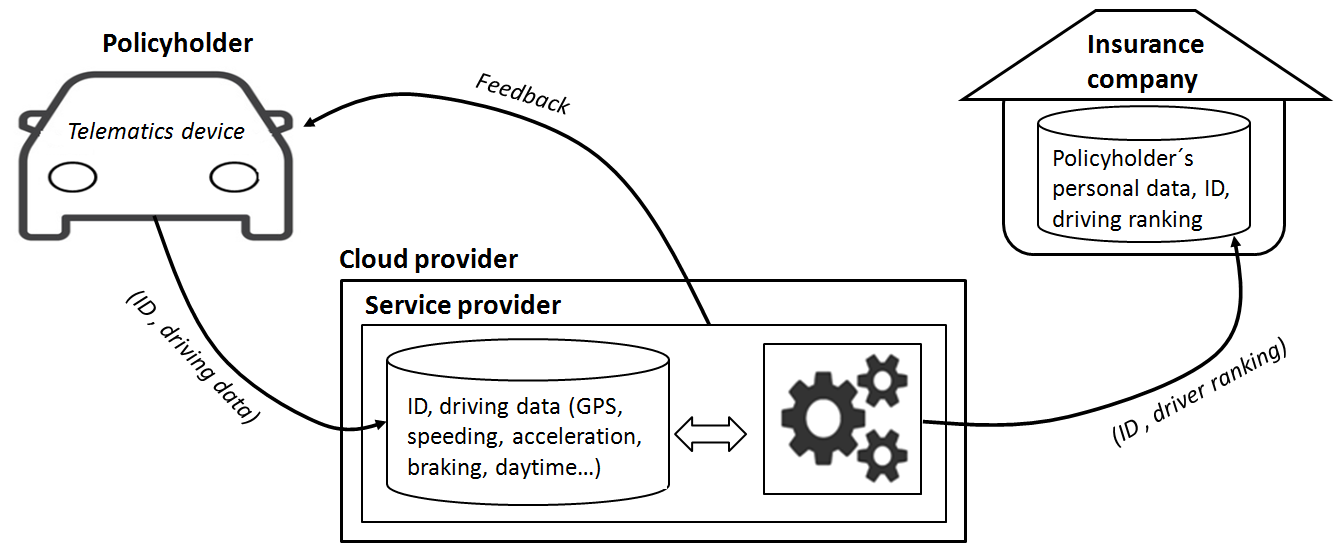}
  \caption{High-level view of usage-based insurance scenario: The
    data is processed by the service
    provider on behalf of the insurance company. Processing is performed by
    a cloud provider running the service provider's software. The
    policyholders receive feedback on their driving habits.}
  \label{fig:appScenario}
\end{figure}

Figure~\ref{fig:appScenario} depicts an abstract view of UBI: The
service provider may actually be the same entity as the insurance
company, but in many business implementations (BonusDrive by Allianz
\citep{allianzUserGuide}, SmartDriver by HUK-Coburg \citep{hukCoburg})
it is a different company. One reason for separation is that insurance
companies do not have the corresponding know-how to compute the driving
ranking. 
Another reason is that the insurance companies want to
mitigate consumers’ privacy concerns by stating that they have no
access to the behavioral data, as it is processed by a third party \citep{bonusDriverInterview}.

Users who process sensitive data in the cloud have the following general security requirements:
\begin{itemize}

\item Confidentiality of data: Policyholders agree that their ranking
  is computed, but they want their individual usage data to remain
  confidential towards the insurance company and the cloud provider. 

\item Confidentiality of code: Service providers want to protect their
  intellectual property from other parties, in particular the
  insurance company and the cloud provider. So the software which is deployed in the cloud should be protected.

\item Integrity of data and code: Insurance company, service provider and the policyholders should have a guarantee that the cloud provider does not change data
  or code in any unauthorized way.

\end{itemize}

On the one hand, users establish a sense of trust in the cloud provider
in practice via contracts over \emph{Service Level Agreements} (SLAs), auditing
certificates and reputation. 

Unfortunately, even with the most refined SLAs the necessity to place
trust in the cloud provider remains. 

On the other hand, numerous technical approaches \cite{univis91762618}
have been proposed to achieve security requirements such as those
above using trusted hardware. For example, \emph{hardware security
  modules} (HSMs) \cite{DBLP:journals/computer/DyerLPSDSW01,utimaco},
i.e., tamper-resistant physical computing devices, can perform secure
and confidential computation of data. Using HSMs, it is possible to
deploy specific software modules, create cryptographic keys and
process data purely within the hardware device. Returning to our UBI
scenario, the HSM can be used to effectively protect the service
provider's data and code from the cloud provider. However, in this
case the necessity to trust a single entity is not avoided, it is
merely shifted from the cloud provider to the trusted
hardware provider. This observation is not specific to HSMs but holds also for
other such technologies such as Intel SGX
\cite{Baumann2014:ShieldingAppCloudHeaven,Schuster2015:TrustworthyDAcloudSGX}.

\subsection{Contributions} 
\label{sec:contribution}

In this paper, we propose a general approach that ensures generic
confidentiality and integrity of cloud service and that avoids the
necessity of having to trust a single entity.

Our approach is based on the combination of two concepts:
\begin{enumerate}
\item \emph{Sealed computation}, an abstract
  technical mechanism to confine the processing of data within a
  tamper-proof hardware container (like a HSM), and
\item a procedural mechanism of mutual checking applying the
  additional role of an \emph{auditing party}, which is necessary to
  check whether the system works as expected, but cannot modify it.
\end{enumerate}
We describe the abstract technical and procedural requirements of
both concepts and argue that they are sufficient to achieve the
generic security properties described above.  In the spirit of work by
\citet{Morris1973:protectionProgramming}, our work is conceptional,
avoiding over-formalization but still providing clear definitions and
evaluating statements.  The main insight is to show how an abstract
hardware mechanism (sealed computation, solely defined by its
requirements) must be utilized in the cloud service such that the
necessity to trust in a single entity is avoided.

Similar to other work
\cite{Schuster2015:TrustworthyDAcloudSGX,Baumann2014:ShieldingAppCloudHeaven},
this paper focuses on integrity and confidentiality properties and
do not consider availability. We use the UBI scenario above
repeatedly as an example to illustrate our exposition, it generalizes to many other scenarios.

\subsection{Outlook} 
\label{sec:outlook}

We first define the concept of sealed computation in
Section \ref{sec:sealedComputation}. Then  
 the system and attacker model is presented in
Section~\ref{sec:system:model}. 
Section~\ref{sec:TheModel:sealedWithAuditor} describes the procedural mechanism applying the role of an auditor.
  
In Section~\ref{sec:SecAnalysis:Discussion} we provide a security
analysis and argue that general security requirements are satisfied unless
two parties act maliciously. Related work is discussed in Section~\ref{sec:related:work}. Finally, Section~\ref{sec:conclusions} concludes the
paper.

\section{Sealed Computation} 
\label{sec:sealedComputation}

While data at rest can, typically, be protected by
encryption, while protecting
 data during
processing commonly is still an interesting problem to solve.  
We introduce a definition of sealed computation using abstract
roles to keep it general, later, these are mapped to the
parties introduced in Section~\ref{sec:system:model}.
 The term \emph{sealed computation} is an abstraction that describes a
well-defined level of protection against such attackers. Intuitively,
this is done by encapsulating the software execution within a physical
piece of hardware. We utilize the notion of sealed computation to
 maintain the integrity and confidentiality requirements of the
system.

\subsection{Definition}
\label{sec:definition}

In sealed computation, a party $A$ provides a physical execution
container $C$ into which a party $B$ may ``seal'' its software. The
container $C$ ensures that the software is running in an unmodified
fashion. Furthermore, $C$ also guarantees that only a restricted set
of interactions with the software are possible through a well-defined
interface. Apart from that, no information is leaked from within $C$
to the outside, not even to $A$ the provider of the container nor the
software provider $B$.

More formally, let a party $A$ provide a physical execution container
$C$ and party $B$ provide a software $M$ which implements some
input/output specification via a well-defined interface. The interface
can be thought of as a description of input/output signals over wires or
the format of incoming or outgoing protocol messages.

\begin{definition}[Sealed Computation]
  \label{def:sealed:computation}
  We say that \emph{$B$ seals $M$ within $C$ provided by $A$} if the
  following technical requirements are met:
  \begin{itemize}

  \item (Sealing) $A$ and $B$ cannot access the code and data of $M$
    after it has been sealed within $C$, apart from changes allowed by
    the interface.
    
  \item (Attestation) As long as $M$ has not terminated and as long as
    $A$ acts honestly, $C$ can provide evidence which proves that $C$
    is running software provided by $B$ in a manner which is unique to
    the sealing instance, i.e., any change of $M$, $C$ or any
    subsequent sealing using the same combination will result in
    different evidence.

  \item (Black-box) Information flow between $M$ and any other party
    (including $A$ and $B$) is restricted by the interface
    specification of $M$, i.e., nothing about the internal state of
    $M$ (code and data) can be learned apart from what is given away
    via the interface.

  \item (Tamper-resistance) Any usage of $M$ that does not satisfy the
    interface specification results in termination of $M$ and the
    destruction of $C$ such that neither code nor data from within $C$
    can be retrieved.

  \end{itemize}   
\end{definition}

Intuitively, the Sealing requirement of sealed computation binds the
execution of a program to a particular hardware environment. The
requirements of Black-box and Tamper-resistance limit access to data and
code only to  
interactions given in the functional
specification of $M$:  Black-box restricts information flow for
expected interactions, while Tamper-resistance does this for
unexpected interactions. 

The Attestation requirement enables external parties to validate the fact
that $M$ has been sealed. It implies that $C$
contains some known unique characteristic that can be validated by
checking the provided evidence. This validation, however, depends on
the correctness of $A$. A common realization of this is for $A$
to embed a secret key within $C$ and allow external parties to
validate its existence by providing the corresponding public key. The
existence of such a unique characteristic implies that it is possible
to establish an authentic and confidential communication channel to
$M$ once sealing has started. 

Similarly, note that $B$ or any user of $M$ still has to rely on $A$
to act honestly because it is not verifiable whether $C$ actually
implements sealed computation. However, \emph{if} $B$ correctly seals
$M$ within $C$ provided by an honest $A$, even $A$ cannot change $M$
afterwards and the tamper-resistance requirement of $C$ protects all
secrets within $M$ that are not accessible via its interface or before
sealing.

\subsection{Confidential Software Deployment}
\label{confDeployment}

The notion of sealed computation is a powerful abstraction that can be
used to describe techniques that protect software also during
deployment. We now argue that the technical requirements of sealed computation
allow to ensure the confidentiality of the code which is sealed.

Intuitively, the idea of confidential software deployment is for $B$
to initially install within the sealed computation a \emph{loader stub}
which is able to load the final user program specified by $B$ into
$C$. Within the sealed computation, this software is decrypted,
installed and then takes over the final interface operations expected
by the users. This loader stub can be part of the sealed computation
mechanism from the start. Since it can be easily added to any
mechanism that satisfies Definition~\ref{def:sealed:computation}, we
did not include it as an additional requirement in that definition.

Observe that $M$ cannot be assumed to remain confidential if $A$ is
untrustworthy. However, \emph{if} $A$ is trustworthy, sealed
computation can be used to run code that remains confidential even
towards $A$.

\section{System and Attacker Model} 
\label{sec:system:model}

\subsection{Participants}
\label{sec:participants:reqs}

For a general cloud-based application system model, our approach
assumes the following main participants - referred to as entities or parties interchangeably:
\begin{enumerate}

\item \emph{Data Prosumer} (DP): The DP is a producer and/or consumer of
  data at the same time, i.e., it produces input data and/or has an
  interest to consume the computed results. 
  The way in which data is processed by the
  application is described by the DP in the form of a
  \emph{functional specification}.

\item \emph{Application Software Provider} (ASP): The ASP develops and
  maintains the analytics software which processes the data in the
  cloud and computes desired results according to the
  functional specification.

\item \emph{Cloud Provider} (CP): The CP provides the cloud service
  which includes the hardware infrastructure, the software,
  and all associated configuration, administration and deployment
  tasks. The CP is also responsible for the 
   security of the system as well as its availability towards the DP.

\item \emph{Auditing Party} (AP): The AP is an independent party that
  helps to ensure the integrity of the hardware and
  software before the system becomes operational. We simply
  refer to the AP as the \emph{auditor}.

\item \emph{Sealed Computation Provider} (SCP): Additional entity to
  be considered is the SCP provides the sealed computation technology.

\end{enumerate} 
To map the sealed computation definition in Section~\ref{sec:sealedComputation} to the UBI scenario, it may help to think of the execution container $C$ being a
specific HSM provided by party $A$ (the SCP), while party $B$ is the service
provider (SP) who wrote software $M$ on behalf of insurance company (DP).

\subsection{User Security Requirements} 
\label{sec:reqs:assumption}

The desired security requirements of the parties are described in more
detail here. Every requirement has a name that is prefixed by the
corresponding participant role.

\begin{definition}[User Security Requirements]
  \label{def:requirements}
  The participants have the following \emph{security requirements}:
   
  \begin{itemize}

  \item (DP-Privacy) The DP requires that data remains confidential to
    any other party, i.e., neither CP, nor ASP, nor AP, nor SCP can
    learn anything about the data.\footnote{While privacy has many
      definitions, here we explicitly use the term Privacy and not
      Confidentiality to emphasize end users' privacy (as individuals)
      against the providers and operators of the system (as
      organizations).}

  \item (DP-Integrity) Results which are obtained from the system by
    the DP are correctly computed on data as provided 
    according to the functional specification. DP-Integrity covers
    data storage and processing integrity.

  \item (ASP-Integrity) The analytics software provided by the ASP is
    executed in an unmodified form within the system. Note that
    ASP-Integrity does not imply DP-Integrity since the latter refers
    also to data.

  \item (ASP-Confidentiality) No other party except the AP is able to
    learn about the analytics software developed by the ASP
    apart from what is described in the functional specification.

  \end{itemize}
\end{definition}

\subsection{Attacker Model} 
\label{sec:attackerModel}

In this section, we formulate the attacker model. First, the ways in which
individual participants may maliciously misbehave are described (the
\emph{local} attacker assumption). Then we define a condition
that restricts the number of parties that may act maliciously (the
\emph{global} attacker assumption). The participants may act as follows:
\begin{itemize}

\item Application Software Provider (ASP): The ASP could provide an
  analytics software that leaks information about the processed data,
  thus violating DP-Privacy.  Also,
  the ASP could violate DP-Integrity by providing software that
  incorrectly computes the results, i.e., computes the results not
  according to the functional specification provided by the DP.

\item Sealed Computation Provider (SCP): The SCP could provide an
  incorrect sealed computation mechanism, i.e., a mechanism that has
  back-doors or vulnerabilities that enable changing code and
  data,  
  thus violating ASP-Integrity or DP-Integrity, or a system that leaks code or data 
  which violates ASP-Confidentiality or DP-Privacy.

\item Cloud Provider (CP): The CP could leak any software that it has
  access to a malicious party, thereby violating
  ASP-Confidentiality. The CP has physical access to the mechanism
  provided by the SCP so it may 
  attempt to access and/or modify data that is stored/processed, thus violating
  ASP-Integrity, DP-Integrity or DP-Privacy.

  We assume, however, that the CP protects its systems from
  interference and misuse by external attackers that are not specific to
  our scenario. Therefore these attacks are excluded from consideration in this work.

\item Auditing Party (AP): During checking, the AP could try to add
  functionality to the system to leak information about the
  processed data and/or the software, thereby
  violating DP-Privacy or ASP-Confidentiality directly. 

\end{itemize}

If any party acts in ways described above we say that this party
\emph{acts maliciously}. A party that does not act maliciously is considered
\emph{honest}.

For reasons of simplicity, the DP is excluded from our attacker
model. Typical misbehavior of the DP can be giving a wrong
functional specification, providing false data or to reveal the received results to any other party. Correct
behavior in this respect cannot be enforced using a trustworthy cloud
service as we envision here. Therefore, the DP is assumed to always be
honest.

The \emph{global} attacker assumption, i.e., a
restriction on the number of parties that may act maliciously is formulated as follows: either the AP or both SCP and ASP are honest. More
precisely, if the identifiers are taken as Boolean predicates of
whether they are acting honestly or not, then the global attacker
assumption is satisfied if the following condition holds:
$$AP \lor (SCP \land ASP)$$
Note that the condition is independent of the actions of the CP, and
that it does not state which party exactly acts maliciously (AP, SCP or ASP).

\subsection{Availability of Remote Attestation}
  
To establish trust, it is often necessary to use mechanisms for
\emph{remote attestation}. Following the terminology of Cocker et
al.~\cite{DBLP:journals/ijisec/CokerGLHMORSSS11}, attestation is the
activity of making a claim to an appraiser about the properties of a
target by supplying evidence which supports that claim.  An attester
is a party performing this activity. The result of an attestation
depends on a mixture of facts that the appraiser can check directly on
the evidence provided by the attester (e.g., cryptographic signatures)
and trust in the attester itself (the mechanism by which the evidence
was generated). Any party being part of a remote attestation has the
requirement that the directly checkable part of the attestation works
as expected. In practice, this means that the used cryptography (e.g.,
digital signatures) is secure and that honest parties protect their 
cryptographic secrets.

\section{Combining Sealed Computation with an Auditor} 
\label{sec:TheModel:sealedWithAuditor}

One application of sealed computation in cloud computing would be for
the CP to offer a mechanism to its ``customers'' DP and ASP to perform
a sealed computation on the provided cloud hardware. In this case SCP
and CP would be the same party. However, note that utilizing sealed
computation alone is not sufficient to ensure the participants'
security requirements because (1) sealed computation does not
guarantee anything before sealing takes place, and (2) the mechanism
of sealed computation cannot be trusted without means to verify its
function.  We will therefore treat CP and SCP as independent parties.

\subsection{The Role of Auditor}

The sealed computation is combined with the role of an auditing party AP to
establish the security requirements described in
Definition~\ref{def:requirements}.  In general, auditors are known to
usually perform independent checks and assess other entities in terms
of service, performance, security and data privacy
and system operations \cite{Habib2012:trustFacilitatorCloud}. We use
the AP to both guarantee the functionality of the sealing mechanism
provided by the SCP and to verify the functionality of the analytic
software provided by the ASP.  Once sealing has taken place, the
mechanism of sealed computation ensures continued trust in the system
without having to interact with the AP anymore.
 
The auditor is \emph{not} allowed to add or modify
functionality in the system.  
This is ensured by
a mutual checking procedure described below. The AP, however, has to
enable a possibility of attestation which is independent of the
SCP. This can be realized by either providing an independent mechanism
or (better) by adequately configuring an attestation technique that is already
presented in the sealed computation technology (e.g., by embedding a
secret within the physical container of sealed computing).

\begin{figure}[!h]
 \centering
  \includegraphics[width=1.0\textwidth]{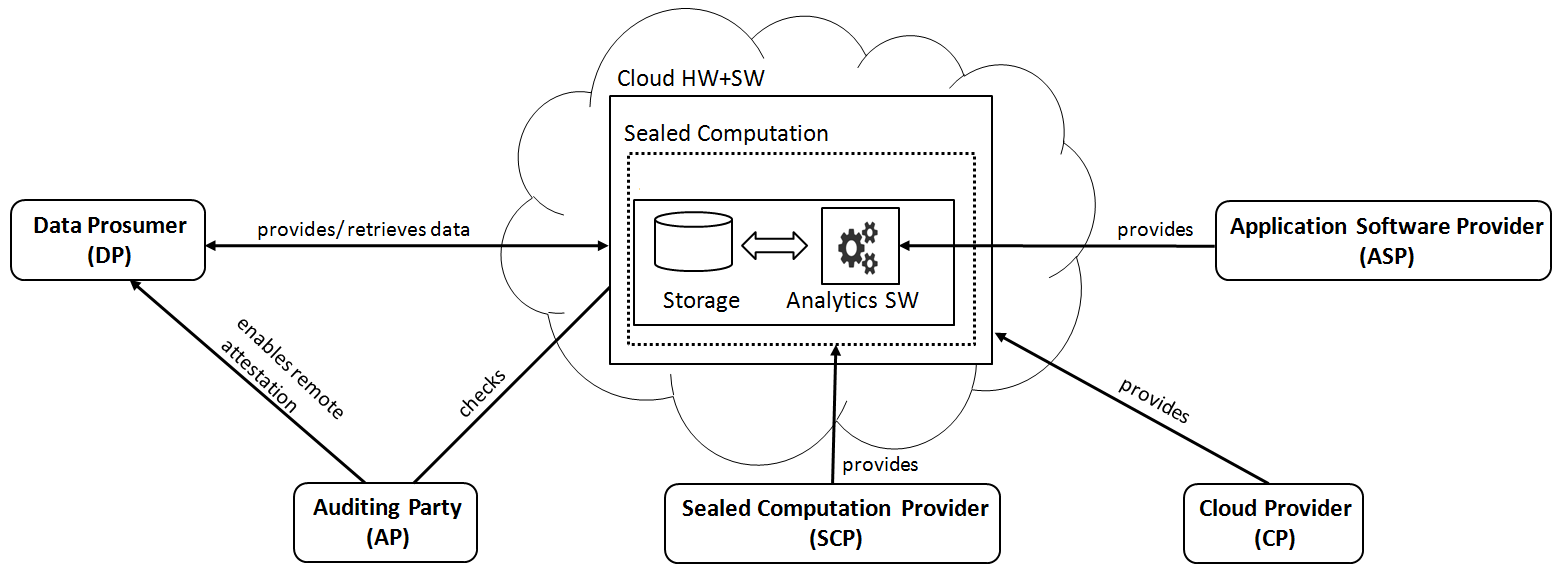}
  \caption{Refined structural model with participants: The ASP
    provides software run within a sealed computation, a mechanism
    provided by the SCP and hosted by the CP. The AP performs an
    independent verification of the analytic software and the sealed
    computation container and enables mechanisms for the DP to
    remotely check its integrity.}
  \label{fig:model}
\end{figure}

Figure \ref{fig:model} illustrates the structural model with the roles
and responsibilities of each participant. The idea is to base
the well-functioning of the system on the assumption that either the
auditor or all parties checked by the auditor are honest during critical phases of system operation. While
commonly the DP had to trust the CP exclusively, it now must rely on
trust \emph{either} in the SCP and ASP \emph{or} the AP (a condition
expressed in our global attacker assumption above).

To illustrate the different roles using our introductory UBI scenario,
the policyholders and the insurance company share the role of the
DP. The insurance company defines the functional specification of the
driver ranking based on which the ASP develops the analytic
software.
The SCP could be a provider of the sealed computation container (like a HSM) and the AP would be a company
like a certified public accountant, that is able to perform code and
security audits on hard- and software.  The SCP is assumed to have
appropriate security mechanisms in place against attacks by parties
not considered above (e.g., hackers and cybercriminals). Regarding
remote attestation, the HSM provides certificates with which
attestation evidence generated by the HSM can be verified
\cite{DBLP:conf/isw/WagnerKE13}.

\subsection{Trust Establishment Procedure}

For simplicity and comprehension of discussion we distinguish the
execution lifetime of the system model into mutually exclusive phases:
the \emph{Checking} phase and \emph{Running} phase. During the
Checking phase, the trust establishment procedure takes place, while
the Running phase begins with the service start-up. During the
Running phase, the DP can upload data and get results and the CP
operates the cloud system.

The exact actions and obligations of the participants and interplay
among each other are described as trust establishment procedure
below. This procedure can be regarded as a form of procedural
requirement which in combination with the technical requirements of
Sealed Computation allows to fulfill the user requirements.

\begin{definition}[Trust Establishment Procedure with Mutual Checking]
  \label{def:checking}
  The participants undergo the following procedure:
  \begin{enumerate}

  \item\label{step:trust:asp} Trust establishment in analytics software:
    \begin{enumerate}
    \item The ASP prepares the analytics software ready to be
      deployed.
    \item The AP verifies whether the analytics software satisfies the
      functional specification and does not leak any information about
      the processed data.
    \item At the same time, the ASP ensures that the AP does not
      change any functionality of the analytics software.
    \item As a result of this procedure, ASP and AP generate public
      evidence to be produced by an attestation mechanism by which it
      can be verified that the checked version of the software is
      running in the sealed computation (e.g., a hash of the binary
      code that can be attested).
    \end{enumerate}

  \item Trust establishment in sealed computation mechanism:

    \begin{enumerate}

    \item Before the sealed computation system is shipped and
      deployed, regardless of the deployment model, the SCP prepares
      the sealed computation mechanism (hardware and software,
      including the possibility for confidential software deployment).

    \item The AP verifies (off-line) the integrity of the sealed
      computation mechanism, i.e., the entire hardware and software
      system. This includes a physical check for the security
      measures, policy compliance, data security and data privacy,
      functional check also of the confidential software deployment
      mechanism.
    
    \item At the same time, the SCP ensures that the AP is not adding
      new functionality during these checks, i.e., that the AP is
      behaving according to the auditing procedure specifications.

    \item The AP and the SCP generate public evidence that enables
      attestation of the sealed computation mechanism, e.g., by
      embedding independent private keys within the sealed computation
      container to which they possess the corresponding public keys.
  
    \end{enumerate}

  \item The sealing mechanism is started in the presence of AP and
    SCP. At this time the auditing procedure ends and both SCP and AP
    can leave the deployment site which is run by the CP.
  \item Using the confidential deployment procedure, the ASP loads the
    code that was checked by the AP in Step~\ref{step:trust:asp}
    above.
  \item The AP and the SCP must be present any time when the system
    and/or the sealed computation mechanism is reset/restarted, is
    under maintenance or shall be changed. In such cases the AP and
    the SCP must re-check the system and both must re-enable the
    attestation mechanism as described in the above procedure.
  \end{enumerate}
\end{definition}
The result of this procedure are two pieces of public evidence that
all parties can use to verify their security requirements:
\begin{itemize}
\item Public evidence provided by AP and SCP that DP, CP and ASP can use to
  verify that an instance of sealed computation is running.
\item Public evidence provided by AP and ASP that can be used to
  verify that a particular software is running within the sealed
  computation.
\end{itemize}

\section{Security Analysis and Discussion} 
\label{sec:SecAnalysis:Discussion}

\subsection{Security Analysis}
\label{sec:securityAnalysis}

To argue that the security requirements from Def.~\ref{def:requirements}
are met, we make the following introductory observation: The
sealed computation mechanism defined in
Def.~\ref{def:sealed:computation} will not be in the Running phase if
the ASP software or the sealed computation mechanism is not correct.

To see this, we make a case distinction based on the global attacker
assumption which states that all parties can act maliciously as long
as the global attacker assumption is satisfied, i.e., either the AP or
both the ASP and the SCP behave honestly. 
There are three possible cases for parties to act maliciously during the checking phase when the trust establishment procedure (Definition~\ref{def:checking}) takes place:
\begin{itemize}

\item The ASP is malicious: If the ASP is malicious, then the AP must
  be honest.  So if the ASP acts maliciously and implements an
  incorrect software then the checking procedure (Step 1.b) mandates
  that the AP checks the software correctness. Since the AP is honest,
  it will detect the incorrectness of software, the check will fail
  and the Running phase will not take place.

\item The SCP is malicious: If the SCP is malicious, then the AP must
  be honest.  So if the SCP is not honest, the sealing container may
  not be implemented correctly. However, the checking procedure (Step
  2.b) requires the AP to check whether the sealed computation
  requirements are met. Since the AP is honest, it will detect
  incorrectness and the Running phase will not be entered.
 
\item The AP is malicious: If the AP is malicious, then the ASP
  \emph{and} the SCP are both honest. In this case, the analytics
  software and the sealed computation mechanism are correct from the
  beginning. Furthermore, the mutual checking procedure (Steps 1.c and
  2.c) requires that both ASP and SCP ensure that the AP does not
  manipulate the functionality of the analytics software or the sealed
  computation mechanism. So if the Running phase is entered, the
  sealed computation mechanism and the analytics software are both
  correct.

\end{itemize}

Therefore, under the attacker
assumption, the establishment procedure guarantees that the system will not enter the Running phase
unless it is working properly as defined in the
specification. 

Subsequently, 
during the Running phase, 
the sealed computation mechanism
(Definition.~\ref{def:sealed:computation}) takes over to guarantee the
desired requirements. To argue for the fulfillment of ASP-Integrity and
DP-Integrity, the Sealing and Tamper-resistance requirements of the
sealed computation ensure that content (data and code) in the sealed
container cannot be improperly modified. Furthermore, the Black-box
requirement restricts information flow such that DP-Privacy and
(assuming confidential deployment) ASP-Confidentiality are
maintained.

\subsection{Discussion}
\label{sec:discussion}

While our results are conceptual, 
they provide a preliminary guideline of building a trustworthy cloud computing service in which cloud customers can trust that cloud providers and operators cannot access their data and code. In essence, sealed computation may not be a brand new concept, as sealed storage was defined by Morris \cite{Morris1973:protectionProgramming}. Whereas, to the best of our knowledge, sealed computation was not formally defined comprehensively before.  
Any computational implementation that satisfies the requirements defined in Definition~\ref{def:sealed:computation} can be considered a sealed computation mechanism. However, in practice, one
may argue that any assumption like the security of cryptography or
requirements like Black-box of any hardware device only hold with a
certain probability, so the guarantees in practice never hold with
100\%.  One may also argue that many parts of the procedures described
in Definition~\ref{def:checking} are also rather hypothetical and
cannot be realized fully in practice.  For example, the AP is assumed
to \emph{perfectly verify} the correctness of the software of the ASP
(in Step 1b) against the functional specification. While software
verification has come a long way, it still is restricted by the size
and complexity of the software system. Another example that appears
far from practice is the statement that the AP can verify the
correctness of the sealed computation container (hardware and
software) provided by the SCP (in Step 2b). It is well-known that the
production of hardware is a very complex process involving lots of
different technologies. The resulting chips are rather non-transparent
and need complex validation equipment to be checked.

Useful insights can be inferred from the proposed approach.
While the AP is one party in our model, in practice it can
consist of multiple independent auditing actors, e.g., different
companies that all check independent parts of the system and mutually
certify the results towards each other. The collection of auditors in
its entirety then forms the AP, meaning also that \emph{all}
``sub-auditors'' must behave correctly for the AP to be regarded as
honest. In practice, these sub-auditors are even often part of the
same company, albeit in different parts that are independent of each
others (like software development and testing departments).

Another highlight is,  it \emph{is} possible to delegate
security enforcement to trusted hardware without having to trust a
single entity. However, during the Checking phase, the AP must be
\emph{continuously present} until the sealed computation container
runs, and it must be possible to establish attestation evidence which
is \emph{independently supported} by the AP and the SCP (for the
sealed computation container) and by the AP and the ASP (for the
analytics software).  These points result from the requirement of
mutual checking, i.e., not only does the AP verify the actions of
ASP/SCP, but also ASP/SCP need to prevent the AP from slipping in new
functionality to software and hardware, a detail which is often
overlooked or (unconvincingly) excluded by the assumption that the AP
is always honest. Being able to embed shared attestation credentials
of mutually untrusted parties in a single trusted hardware container
is a feature which is --- at least to our knowledge --- not supported
by any currently available trusted computing mechanism
\cite{univis91762618}.

So overall, the proposed approach presents an idealized version of system
construction and deployment processes which can serve as an orientation
for practice towards achieving a trustworthy service.

\section{Related Work}
\label{sec:related:work}

Privacy is a major factor in trusting data and computation outsourcing, such as in a cloud-based application. Hence trust establishment has been discussed in the context of cloud from different perspectives in the literature, we distinguish them into technical and non-technical trust enhancement approaches. \citet{Georgiopoulou2017:LitRevTrust} reviewed a number of trust models for cloud computing trying to provide a gap analysis in the literature. However, the review considered only a very limited set of models.

Non-technical approaches have been developed and used ranging from SLAs and recommendations for security architecture, risk management and operational teams. For example,  \citet{Alhanahnah2017:cloudTrustFWandTaxonomy} studied a trust evaluation framework to allow cloud customers to choose among set of cloud providers based on trust levels. The authors distinguished trust factors into two sets: SLA-based and non-SLA factors based on the provider's reputation and even financial status.

 \citet{Rizvi2014:centralizedTrustModel} utilized the auditor role to provide an objective trust baseline assessment to enable clients to decide between CP candidates. The proposal delegates the trust assessment to an auditor to calculate trust values. So that clients who need to choose between CPs request the trust values from the auditor based on required service. The auditor role, we present, is not the same as the third party role in these works as shown in the trust establishment procedure~\ref{def:checking}.

Hence the common trust management model in the Web relies on the binding a domain name and a public key, is not enough for privacy in cloud computing. 
A number of solutions were presented to enforce trust via technical means that ensure the privacy of the users data. 
\citet{Santos2012:policySealed} employed attribute-base encryption to provide a policy enforcement protocol based on Trusted Platform Module (TPM) abstraction. Similarly, \citet{Li2010:trustedComputingAuditor} proposed a model to support security duty separation in multi-tenant IaaS cloud between CP and customers based on TPM and they added the auditor role optionally. Moreover,~\citet{Cheng2010:SealedforTrustCloud} proposed to build an architecture for IaaS model
to give the clients trust to deploy their VMs, 
 that provides sealed storage and relies on remote attestation.   
 
These models ~\cite{Santos2012:policySealed, Li2010:trustedComputingAuditor, Cheng2010:SealedforTrustCloud} were designed for Platform as a Service (PaaS) and Infrastructure as a Service (IaaS) cloud models that require less security responsibilities on the CP~\cite{alliance2011:securityGuidanceCloud} as they are shared with the customers, 
 while SaaS model requires more responsibilities from CP~\cite{Rizvi2014:centralizedTrustModel}.

 A trustworthy and privacy-preserving cloud may be addressed by the use of cryptographic techniques such as fully homomorphic encryption (FHE)~\cite{Gentry2009:hommorphicIdealLattices}. However, it is still inefficient for most computations~\cite{Schuster2015:TrustworthyDAcloudSGX}.
 Similarly in verifiable computing \cite{ Parno2013:practicalVerifiableComp},
  it was designed to enable result correctness verification but has not shown support for general purpose cloud computing yet.

\section{Conclusions and Future Work} 
\label{sec:conclusions}

We introduced the sealed
computation concept and proposed a mutual checking procedure with an auditor role during
setup time to provide an increased level of security and trust in
cloud scenarios. The sealed computation concept abstracts from
trusted hardware technologies like HSMs, the auditor is an
abstraction of policies and procedures that increase trust in a single
party.

We believe that the abstract system model using the auditor as an
additional role is a good approach for medium-size and large cloud
deployments instead of running their own private cloud. While the existence of the role of auditor may be
intuitive, on the one hand, it is not clear whether the concept is really
\emph{necessary}, i.e., whether any technique that distributes trust can simulate the auditor as described above. On the other hand, practical methods for auditing could be investigated. Furthermore, we wish to attempt more rigid formalization for the attestation verification. 

\section*{Acknowledgments}

The authors would like to thank Nico D\"ottling, Johannes G\"otzfried, Tilo M\"uller and Hubert J\"ager for hints and useful comments on earlier versions of this paper. This research is conducted under and supported by the ``Privacy\&Us'' Innovative Training Network (EU H2020 MSCA ITN, grant agreement No.675730).

\bibliography{references}

\end{document}